\documentclass[10pt,showpacs,showkeys,preprint]{revtex4-1}
\usepackage{bm}
\usepackage{amssymb}
\usepackage{amsmath}
\usepackage{graphicx}
\usepackage{multirow}
\usepackage{MnSymbol}
\usepackage{wasysym}
\usepackage{stmaryrd}
\usepackage[outline]{contour}
\usepackage[dvipsnames]{xcolor}
\usepackage{float}
\usepackage[mathlines]{lineno}

\definecolor{olive}{RGB}{182,187,37}
\begin{document}

\title{The atomic ionization, capture, and stopping cross sections by
multicharged ions satisfy the Benford law}
\author{J. E. Miraglia and M. S. Gravielle}
\date{\today }

\begin{abstract}
The applicability of the Benford law for different data sets of atomic cross
sections by ion impact is studied. We find that the data sets corresponding
to theoretical ionization and capture cross sections of neutral targets by
multicharged ions satisfy quite well the Benford law, not only for the first
digits but also within a given order of magnitude. Experimental stopping
power values from the International Atomic Energy Agency data base were also
scrutinized, but in this case the Benford conformity was not so satisfactory
due to its small rank width. In all cases, errors, densities of prime
numbers, theorems of Nigrini and Pinkham are evaluated and discussed.
\end{abstract}

\maketitle

\affiliation{Instituto de Astronom\'{\i}a y F\'{\i}sica del Espacio. Consejo Nacional de
Investigaciones Cient\'{\i}ficas y T\'{e}cnicas}
\affiliation{Departamento de F\'{\i}sica. Facultad de Ciencias Exactas y Naturales.
Universidad de Buenos Aires. \\
Casilla de Correo 67, Sucursal 28, {C1428EGA} Buenos Aires, Argentina.}

\section{Introduction}

The Benford's Law (BL) was originally proposed by the astronomer Newcomb\ in
1881 \cite{Newcomb1881} and formulated \ much later by Benford in 1938 \cite%
{Benford1938}.\ For a given big data set $\{\eta \}$, both man-made or from
nature$,\ $the Benford distribution states that the frequency of the first
significant digit $\eta _{1}$ follows the following law%
\begin{equation}
P_{\eta _{1}}^{B}=\log _{10}(1+1/\eta _{1}),  \label{10}
\end{equation}%
where $P_{\eta _{1}}^{B}$ is the Benford probability of occurring $\eta
_{1}\ $(not considering the number of zeros on the left of the numbers).
From a mathematical point of view \ \cite{Berger2011}, to identify the
digits of a given real number, one can define an operator corresponding to
the first $(j=1)$, second\ $(j=2)$ and higher digits$\ (j>2)$ of a real
number $\eta $ as\ $\widehat{D}_{j}(\eta )=\eta _{j}$. For example if $\eta
^{\ast }=0.0234$, then $\widehat{D}_{1}(\eta ^{\ast })=\eta _{1}=2,\ \
\widehat{D}_{2}(\eta ^{\ast })=\eta _{2}=3,\ $and$\ \widehat{D}_{3}(\eta
^{\ast })=\eta _{3}=4.$ The numbers formed by the first two digits is here
denoted as$\ \{\eta _{1},\eta _{2}\},\ ($in our example $\{\eta _{1},\eta
_{2}\}=23)$ and so on. The BL can be extended for all decimal digits: for
example for the first two digits, it holds%
\begin{equation}
P_{\{\eta _{1},\eta _{2}\}}^{B}=\log _{10}\left( 1+\frac{1}{10\eta _{1}+\eta
_{2}}\right) .  \label{20}
\end{equation}%
Like any probability, the closure relation is normalized,%
\begin{equation}
\sum\limits_{\eta _{1=1}}^{9}P_{\eta _{1}}^{B}=\sum\limits_{\{\eta _{1},\eta
_{2}\}=10}^{99}P_{\{\eta _{1},\eta _{2}\}}^{B}=1.  \label{25}
\end{equation}%
In addition, if we are interested in the order of magnitude $M,$ we need to
use the operator $\widehat{M}$,\ so that $\widehat{M}$ ($\eta $)$=M.$ In our
example, $\widehat{M}(\eta ^{\ast })=-2.$

For the first digit, $\eta _{1}\in \lbrack 1,9]$, the BL states that the
number $\eta _{1}=$1 appears in more than 30\% of the cases, while$\ \eta
_{1}=$9 in less than 5\%. But the probability of appearance of the second
digit,$\ $%
\begin{equation}
P_{\eta _{2}}^{B}=\Sigma _{\eta _{1}}P_{\{\eta _{1},\eta _{2}\}}^{B},
\label{255}
\end{equation}%
varies from $P_{\eta _{2}=0}^{B}=0.1196$ to$\ P_{\eta _{2}=9}^{B}=0.0850,$
tending to the \textit{uniform} distribution $P_{\eta _{m}}^{B}\rightarrow
1/10,$ as $m\rightarrow \infty .$

If we consider two-digit integer numbers, so $\{\eta _{1},\eta _{2}\}\in
\lbrack 10,99],$ the probability of appearing $10$ is $P_{\{\eta _{1},\eta
_{2}\}=10}^{B}=0.047,$ while the one of $99$ is $P_{\{\eta _{1},\eta
_{2}\}=99}^{B}=$0.004. \ A list of occurring of a given digit in the first
second and third positions can be found in several articles, for example in
Refs.\cite{Nigrini1999,Nigrini2012}.

At first sight, these results are against the human perception, which would
tend to assume a randomness. For a single digit,$\ $one would guess a
\textit{uniform} distribution, that is, $P_{\eta _{1}}^{U}=$ 1/9 for any
digit, and for\ two digits, $\ P_{\{\eta _{1},\eta _{2}\}}^{U}=1/90,$ and so
on. This is precisely what makes interesting the BL: if the numbers are
naively forged by the human mind, for example, the BL will no longer be
satisfied.

It is important to note that not all data sets follow the BL, but just the
ones resulting of the product of multiple independent factors that produce a
probability covering several orders of magnitudes when they are plotted in a
lognormal scale.

There is a huge amount of applications, like fight against tax fraud by
detecting manipulation as\ anomalies, stock exchange data, corporate
disbursements, analysis of sales figures, demographics and scientific data,
etc. \cite{Nigrini1999,Nigrini2012}. The application of the BL crosses a
huge range of disciplines, including physical sciences. Also astrophysical
data like the exoplanet masses, pulsars rotation frequencies, $\gamma $-ray
source fluxes and fundamental physics constants satisfy the BL \cite%
{Sambridge2011,Alexopuolos2014,Miraglia2021}. A complete list of articles on
BL can be found in Ref. \cite{Beebe2020}.

\section{The atomic data sets}

The values $\eta \ $that we are going to scrutinize under the BL correspond
to three different atomic collision data sets, namely: ionization from the
subshell $nl\ $of neutral atoms by impact of multicharged ions, electron
capture from hydrogen to the subshell $nl$ of bare projectiles, and
experimental stopping power cross sections by multicharged ions moving in
gases. The numbers $n\ $and $l$ denote the principal and orbital quantum
numbers, respectively, of the atomic subshell. Details of these data sets
will be provided in the next subsections.

\subsubsection{ Ionization cross section data set.}

During the last years we have been calculating $nl$-ionization cross
sections by multicharged ions on different neutral atoms. Calculation were
carried out with the continuum distorted wave-eikonal initial state
(CDW-EIS) theory. All the results were numerically performed as accurate as
possible and published in Ref.\cite{Miraglia2019}. \ The main aim of such
calculations was the evaluation of molecular ionization cross sections
within a stoichiometric model \cite{Mendez2020}.

Theoretical ionization cross sections for the following systems:
antiprotons, H$^{+}$, He$^{2+}$, Be$^{4+}$, C$^{6+}$ and O$^{8+}$ impinging
on H, He, Li, Be, B, C, N, O, F, Ne, P, S and Ar neutral atoms for impact
energies ranging from 100 to 10000 kev/amu were put together. We gather 2808
results with three significant figures covering about ten orders of
magnitude. All these values were expressed in atomic units forming a data
set that will be noted, for short as $\eta =I=I(nl).$ We also explore the
reduced set of total ionization cross sections, $\eta =X=\Sigma
_{nl}o_{nl}I(nl),\ $where\ \ $o_{nl}\ $is\ the occupation number of the $nl$%
-subshell.

\subsubsection{Electron capture cross section data set}

We consider capture cross sections from hydrogen to the subshell $nl$ of the
impinging bare ions calculated with the eikonal impulse (EI) approximation.
Results for capture by H$^{+}$ and He$^{2+}$ projectiles to the principal
quantum numbers $n=$1,2,3, and 4, and by\ Li$^{3+},\ $Be$^{4+},\ $B$^{5+},$ C%
$^{6+},$ N$^{7+}$ and O$^{8+}$ ions to the principal quantum numbers $n=$%
1....8 , both for orbital momentum numbers $l=0,...n-1,$ are analyzed. . In
all the cases, 13 impact energies \ ranging from 25 to 1220 keV/amu. were
considered\ \cite{Jorge2015}. \ We totalize a set of 3275 values which ,will
be here noted, for short as $\eta =C=C(nl).\ $Also\ the smaller set of
capture\ cross sections to a given principal number $n,$ defined as $\eta
=T=T(n)=\dsum\limits_{l}C(nl),$ will be explored.

\subsubsection{Stopping power data set}

We use data of experimental stopping power as stored by the web site of the
International Atomic Energy Agency (IAEA) \cite{iaea}. \ We collected all
the values corresponding to heavy ion impact on neutral gas targets. We
bring together 4118 values from 52 files of the web site\ for the following
colliding systems: H$^{+}$,\ He$^{++}$,\ Li$^{3+}$,\ \ N$^{9+}$,\ Cu$^{29+}$%
and\ \ Kr$^{36+}$\ impinging on$\ $H$_{2},$\ He,\ N$_{2}$,\ O$_{2}$,~Ne,\
Ar,\ Kr, and\ Xe.\ All the values were standardized to atomic units from the
practical units 10$^{-15}$eVcm$^{2}$ and Mev\ cm/mg, as published in the
site. Furthermore, the targets were described as they are found in nature,
that is, for dimers, such as H$_{2},$ N$_{2}$,\ and O$_{2}$, the results of
the IAEA were multiplied by 2 because they are normalized to the number of
atoms. Note that this data set represents a great challenge for the BL
because it contains several experiments differing each other, specially in
the intermediate and low energy regions where we may have different first
digits for the same collision parameters. This set will be noted as $\eta =S$

Summarizing, the following sets will be considered%
\begin{equation}
\eta =\left\{
\begin{array}{ll}
I=I(nl) & \text{Ionization \ cross section from the }nl\text{ level.} \\
X=\dsum\limits_{nl}I(nl),\ \ \  & \text{Total ionization cross section.} \\
C=C(nl) & \text{Capture\ cross section to the }nl\text{ level.} \\
T=T(n)=\dsum\limits_{l}C(nl),\ \  & \text{Total capture\ cross section to
the }n\text{ level.} \\
S & \text{Stopping power cross section.}%
\end{array}%
\right.   \label{26}
\end{equation}%
These sets have no node, and they are unimodal containing only positive
quantities, which make them good candidates to check if the BL is satisfied.
To study any set of numerical results under the Benford scheme, one
important point to bear in mind is the role of the last digit. In a
numerical calculation, the last digit is generally worked through a rounding
up process, while the last digit of a Benford distribution is understood via
a truncation process. Therefore in our study of the BL the third (last)
digit will be out of discussion.

\subsection{The width and the density of points}

A first requirement to inspect the BL is that the data set is expected to be
evenly distributed in a logarithmic scale. It \ means that the numbers, when
ranked from smallest to largest, can be approximated by a linear form in a
logarithmic scale, that is,
\begin{equation}
Log_{10}\eta \simeq a+bj,  \label{27}
\end{equation}%
where $\eta $ represents a given value of the set and$\ j$ indicates its
order index in the data set, i.e. $j\in \lbrack 1,J$] and $J$ is the total
amount of values of the set.\ In all the cases, the actual values of $%
Log_{10}\eta \ $do not fall on a straight line from start to finish;\ it is
reasonably straight in the middle, but it is curvy in the tails. Following
to Nigrini \cite{Nigrini2012}, the width $\Delta $ of a set of values can be
estimated by the difference between the extremes of the linear fitting,
\begin{equation}
\Delta =(a+bJ)-(a+b)=b(J-1)\simeq bJ.  \label{29}
\end{equation}%
For the five sets studied in this article, the values of $\Delta $ are shown
in the Table, and displayed in Figure 1. The width $\Delta $ represents the
effective number of orders of magnitudes that cover the data set. The BL
requires $\Delta \gg 2$ to sample the first digits at least twice. One
expects the larger the width, the more robust the prediction of the BL.

For the data sets I and C, the $\Delta $ values are 5.34 and 9.31,
respectively, being large enough to warranty a good spread of numbers.
Instead for the stopping $S$, $\ \Delta $=2.32, being a small value which
originates critical problems of borders, as we will see. A similar problem
arises if we reduce the number of data. For example, by considering the
total ionization cross section set $X$, the total number of data is reduced
to $J=$912, with values covering just two orders of magnitude. \

In order to measure this effect, we can define a new important magnitude:
the density of points $\delta =J/\Delta \simeq $ $1/b$. From our experience,
we find that it is required $\delta \gg $ $200$ to have a reasonable
occurrence of the first digit only. For total capture cross sections to
different $n-$shells,\ represented by the set $T$, the resulting range $%
\Delta \ $ still covers seven orders of magnitude, but the amount of values
reduces dramatically to $J=755$, given rise to \ $\delta \sim 10^{2}$ which
is a very small density. Hence, we are in a presence of different
situations, for $X$,\ $\Delta $ is substantially reduced, while for $T$, $%
\delta $ is too small. Results are displayed in the Table for all the cases.

\subsection{The degree of "Benfordness"}

At this stage one should develop a tool to quantify how good the data
conform the BL, or in other words, to quantify the "Benfordness" of a \
given data set under study. A traditional approach is to use the chi-squared
statistic. However, this test is not useful for very large data sets since
for\ a large value of $J,$ the calculated chi-square will generally be
higher than the critical value, leading us to conclude that the data set
does not conform the BL \cite{Nigrini2012}. A specific method to measure the
Benfordness of the data is the mean absolute deviation (MAD) test. For the
first digit the MAD test is defined in percentages as \cite{Nigrini2012}%
\begin{equation}
\varepsilon _{1}^{B}=\frac{1}{9}\sum\limits_{\eta _{1}=1}^{9}\left\vert
P_{\eta _{1}}-P_{\eta _{1}}^{B}\right\vert \times 100,  \label{31}
\end{equation}%
where $P_{\eta _{1}}$ is the actual frequency of the first digit of the
magnitude $\eta $\ and $P_{\eta _{1}}^{B}$ is the corresponding prediction
of the BL, as given by Eq.(\ref{10}). Nigrini \cite{Nigrini2012} published
some guidelines, based on his personal experience, to qualify the
Benfordness according with the value of $\varepsilon _{1}^{B}$ as it follows
\begin{equation}
\left\{
\begin{array}{ll}
0.0\%\lesssim \varepsilon _{1}^{B}\lesssim 0.6\%,\ \ \  & \text{\textit{%
close conformity,}} \\
0.6\%\lesssim \varepsilon _{1}^{B}\lesssim 1.2\%, & \text{\textit{acceptable
conformity,}} \\
1.2\%\lesssim \varepsilon _{1}^{B}\lesssim 1.5\%, & \text{\textit{marginal
conformity.}}%
\end{array}%
\right. .  \label{41}
\end{equation}%
If we rule out the rigor of the accountancy to detect frauds, we would be
very satisfied if we can comply with these figures. Just for comparison we
could define equivalently a MAD error corresponding to the uniform
distribution%
\begin{equation}
\varepsilon _{1}^{U}=\frac{1}{9}\sum\limits_{\eta _{1}=1}^{9}\left\vert
P_{\eta _{1}}-P_{\eta _{1}}^{U}\right\vert \times 100,  \label{43}
\end{equation}%
where $P_{\eta _{1}}^{U}=$1/9 is the uniform (random) prediction. \ One
would expect that the ratio%
\begin{equation}
\rho _{1}^{B}=\frac{\varepsilon _{1}^{B}}{\varepsilon _{1}^{U}}  \label{431}
\end{equation}%
\ to be $\rho _{1}^{B}\ll 1$ to indicate that we are in a presence of a
Benford distribution.

As far as the first digit is concerned there is an interesting alternative
test, introduced by Nigrini\ \cite{Nigrini2012} to diagnose a Benford
distribution.\ Consider an any-digit set of a given magnitude $\eta =\{\eta
_{1},\eta _{2},...\}\ $\ and calculate%
\begin{equation}
N_{\mu _{1}}=\frac{\sum\limits_{\eta }\{\eta _{1},\eta _{2},...\}\delta
_{\eta _{1},\mu _{1}}}{\sum\limits_{\eta }\{\eta _{1},\eta _{2},...\}},
\label{44}
\end{equation}
with $\mu _{1}\in \lbrack 1,9].\ $The second theorem of Nigrini \cite%
{Nigrini2012} states that the Benford distribution produces $N_{\mu
_{1}}\rightarrow N_{\mu _{1}}^{B}=1/9.$\ Therefore, this is another
independent strategy to check whether the set under study satisfies the BL.
This new parameter\ allows us to introduces an equivalent Nigrini MAD\ error
$\varepsilon _{1}^{N}$ to assert the Benfordness of the set under study, as%
\begin{equation}
\varepsilon _{1}^{N}=\frac{1}{9}\sum\limits_{\eta _{1}=1}^{9}\left\vert
N_{\eta _{1}}-N_{\eta _{1}}^{B}\right\vert \times 100.  \label{46}
\end{equation}

The analysis of the second digit demands a more fine attention because\ $%
P_{\eta _{2}}^{B}$ is close to randomness $P_{\eta _{2}}^{U}=$1/10 and a
differentiation is required. To this end, we find convenient to define the
ratio of MAD errors%
\begin{equation}
\rho _{2}^{B}=\frac{\varepsilon _{2}^{B}}{\varepsilon _{2}^{U}}=\frac{\frac{1%
}{10}\sum\limits_{\eta _{2}=0}^{9}\left\vert P_{\eta _{2}}-P_{\eta
_{2}}^{B}\right\vert }{\frac{1}{10}\sum\limits_{\eta _{2}=0}^{9}\left\vert
P_{\eta _{2}}-P_{\eta _{2}}^{B}\right\vert },  \label{45}
\end{equation}%
which gives an indication of the goodness of the second digit to conform the
Benford distribution. In similar fashion with $\rho _{1}^{B},$ we can state:
\ if $\rho _{2}^{B}>1,$ the second digit of the set is ruled by randomness,
while $\rho _{2}^{B}<1$ indicates that the set is ruled by Benford. Then,
the smaller $\rho _{2}^{B},$ the closer to the Benford distribution. Values
of\ $\varepsilon _{1}^{B},\ \varepsilon _{1}^{U},\ \varepsilon _{1}^{N},$ $%
\rho _{2}^{B}~$and $\ \rho _{2}^{B},$ for all the sets are shown in the
Table.

\section{Results}

\subsubsection{Ionization cross sections}

We start analyzing the probability of occurring the first figure of \ $nl$%
-ionization data, denoted as $P_{I_{1}}$ in Fig. 2(a). We compare $P_{I_{1}}$
with the Benford\ prediction $P_{I_{1}}^{B}$ given by Eq.(\ref{10}) and
indicated as a histogram in light grey. The agreement is very good and in
consequence the MAD\ error is very small,$\ \varepsilon _{1}^{B}=0.52\%$\
(see Table) which means that, according to (\ref{41}), we can certify the
agreement as\textit{\ close conformity}. The second theorem of Nigrini
produces a very similar error, $\varepsilon _{1}^{N}=0.62\%$, which stands
as an alternative criterion to assess the Benfordness using the same
categorization of (\ref{41}).

We can go further by studying the probability to have a given second digit $%
I_{2}\in \lbrack 0,9],$ denoted as $P_{I_{2}},$ shown in Fig 2(b). Again,
the agreement with the Benford prediction is quite good.

To get deeper the analysis, we should proceed to calculate the two-digit
probability $P_{\{I_{1},I_{2}\}}.$ But instead, we prefer to introduce a
novel criterion based on the prediction of the density of two-digit prime
numbers $\pi _{2},$ which is a very sensitive and sharp value, reduced to
the range [10-99]. This represents a new parameter that subvert human
thinking. One would tend to think that\ $\pi _{2}$ equals that of the
uniform distribution,\ $\pi _{2}^{U}=21/90=0.233$.\ But this is not true for
a Benford distribution. By using Eq.(\ref{20}) one can easily obtain $\pi
_{2}^{B}=0.266$. Our \{$I_{1},I_{2}$\}$\ $\ data produce $\pi _{2}=0.262\ $%
in close agreement with the BL prediction. Therefore, we can conclude
categorically that our $nl$-ionization cross section data set satisfies very
well the BL.

We are also interested in studying the BL within a given order of magnitude.
In Fig. 2(c) we plot the data corresponding to the occurrence of a single
digit within a given order of magnitude,\ that is $\ I_{1M}=I_{1}\times
10^{M}$ . \ Even though the results for $\ I_{1M}$ spread along 9 orders of
magnitudes, from $10^{-8}$ to $10^{0}$, it becomes evident that the BL still
applies within each order of magnitude. To visualize this behavior more
clearly, we first define a top function for a given order of magnitude $M$ as%
\begin{equation}
T_{M}=\sum\nolimits_{I_{1}}P_{I_{1M}},  \label{130}
\end{equation}%
which plays the role of a closure relation within the range of magnitude $M$%
. \ Hence the probability $P_{I_{1M}}$ can be estimated by simply \textit{%
Benfordizing} the top function, i.e.%
\begin{equation}
P_{I_{1M}}^{T}\simeq \log _{10}(1+1/I_{1})T_{M}\ ,  \label{140}
\end{equation}%
which is plotted in in Fig. 2(c). From this figure one can observe that $%
P_{I_{1}M}^{T}$ guides quite well the data values, indicating that all the
substantial information can be reduced to $M$ values of the top function $%
T_{M}.$

On the other hand, the total cross section set $X$\ is another matter
because it is spread in a little more than two orders of magnitude (see the
Table).\ But we can still verify that $X$ follows the first-digit BL with an
error of $\varepsilon _{1}^{B}=1.2\%,$ which means that we can certify the
agreement as \textit{acceptable conformity}.

\subsubsection{Capture cross sections}

Fig. 3(a) shows the probability of occurring the first figure of the $nl$%
-capture cross section data set$\ P_{C_{1}}$ which is compared with the
Benford\ prediction given by Eq.(\ref{10}). The agreement is good and \ the
MAD\ error is small, $\ \varepsilon _{1}^{B}=0.57\%$\ , complying with the\
\textit{close conformity }according with the categorization (\ref{41}). The
second digit does not look so well but it is better than the uniform
distribution, with $\rho _{2}^{B}$<1 (see the Table).
Moreover, the \ distribution along ten orders of magnitudes follows
reasonably well the "Benfordization" of the top probability shown in light
grey in Fig. 3(c).

Instead for the capture cross section to the $n-$state data set $T$, \ the
first digit\ probability degrades to \textit{acceptable conformity }(see
Table), while its\ second digit does not present any bias to Benford nor to
uniform because $\rho _{2}^{B}=1.03.$

\subsubsection{Stopping power cross sections}

This is a case in which not only the width $\Delta \ $is small, but also we
are dealing with a large variety of experimental data from different
laboratories. As shown in Fig. 4(a) the first digit probability $P_{S_{1}}$
resembles to the histogram of Benford, but the MAD error is high, $%
\varepsilon _{1}^{B}=1.52\%,$ \ which hardly qualifies within the category
of \textit{marginal conformity,} as classified in (\ref{41}). Nonetheless,
the shape is definitively more close to the BL than to the uniform
distribution ($\varepsilon _{1}^{U}=5.3\%$ ). The second digit distribution
(Fig. 4(b)) does not differ a lot from the one of Benford because $\rho
_{2}^{B}=0.46<1$. \ On the contrary, the magnitude $P_{S_{1M}}$ does not
follows the logarithmic structure within each order of magnitude as in the
previous cases, as shown in Fig. 4(c).

\subsection{Two numerical experiences}

\subsubsection{Testing the universal scaling of Pinkham.}

One of the most extraordinary property of the BL is its scale invariance, to
the point that it is possible to obtain it mathematically by just invoking
this property, as discovered by Pinkham\ \cite{Pinkham1961}. From the
physics point of view, this is evident because BL should be independent on
the units that we use. Ideally, as the Benford distribution is based on the
logarithmic function, any multiplicative factor will just shift the
distribution keeping intact the occurrences of the digits. This is the case
if we have a big width $\Delta $. However, there is a limitation when the
width $\Delta $ is small, like in the case of the stopping power cross
sections. This limitation surfaces clearly due to the border effects in a
limited width. As $\Delta $ is small, the reduction or enhancement of the
probabilities corresponding to some digits will be transferred to other
digits when a different unit is used.

In our analysis, we use atomic units for ionization cross sections, but we
could have used any other magnitude, that is cm$^{2}$ or whatever we choose.
For example, if we transform the set of $nl-$ionization\ values \ $I$ in cm$%
^{2}$ we need to multiply by 2.800$\times 10^{-17}$. This change of scale
produces $\varepsilon _{1}^{B}=0.43$\% which lightly differs from from the
error 0.52\% obtained with atomic units, i.e., a difference of 0.1\%. It is
important to note that the order of magnitude of the units is irrelevant
since it just replicate the same digits. But if we built the set of stopping
power cross sections in \ 10$^{-15}$eVcm$^{2}$, we should multiply the set
by 1.213, finding $\varepsilon _{1}^{B}=1.9$\% \ which is substantially
greater than the error $1.5\%$ obtained when expressed in atomic units (that
is a difference of 0.4\%). This difference is a consequence of the border
effects due to the small width $\Delta $. Note that in units of 10$^{-15}$%
eVcm$^{2},$ the stopping power set no longer classify as \textit{marginal
conformity.}

\subsubsection{The whole data set}

We essay a daring experience by gathering all the data, $nl$-ionization, $nl$%
-capture and stopping power cross sections (all expressed in atomic units).
We totalize 10201 values in only one numerical universe. We found that this
big set of values satisfies reasonable well the BL with $\varepsilon
_{1}^{B}=0.67$\% and $\rho _{2}^{B}=0.33$\%.\ However it does not represent
an overall improvement of the BL conformity. The explanation is related to
the stopping power set which adds 4118 values in a short width, introducing
a high density that disturbs the required even distribution in the
logarithmic scale. This behaviour demonstrate that a large number of values
not necessarily improves the Benford distribution, but the stability of the
density of points.

\section{Conclusion}

We have studied three different sets of atomic data with the BL, having
three different qualifications, according to the Nigrini scheme: \textit{%
close, acceptable } and \textit{marginal conformities} corresponding to the $%
nl $-ionization, $nl$-capture and experimental stopping power cross sections
of multicharged ions on gases, respectively. This findings allows us to
conclude that any atomic-collision data set having the appropriate width $%
\Delta \ $and density of points $\delta $ will satisfy the Bendford
distribution. Furthermore, the parameters here introduced to quantify the
degree of conformity of the BL, like the MAD errors, could be used as a
useful tools to check the quality of any atomic data set and detect
systematic experimental errors or theoretical biases.

The authors acknowledge the financial support from the following
institutions of Argentina: Consejo Nacional de Investigaciones Cient\'{\i}%
ficas y T\'{e}cnicas (CONICET), Agencia Nacional de Promoci\'{o}n Cient\'{\i}%
fica y T\'{e}cnol\'{o}gica (ANPCyT), and Universidad of Buenos Aires.


\begin{table}[htbp]
\caption{MAD errors, ratios, widths and densities as defined in the tex for
the data sets given in (5) }%
\begin{tabular}{|l|c|c|c|c|c|}
\hline
$\eta$ & \ \ \ $I$ \ \ \  & \ \ $X$ \ \  & \ \ $C$ \ \  & \ \ $T$ \ \  & \ \
$S$ \ \  \\ \hline
J & 2808 & 925 & 3275 & 755 & 4118 \\
$\Delta$ & 5.34 & 2.78 & 9.31 & 6.96 & 2.32 \\
$\delta$ & 413 & 331 & 351 & 108 & 1778 \\
$\varepsilon^{B}_1 $ (\%) & 0.52 & 1.02 & 0.46 & 0.76 & 1.52 \\
$\varepsilon^{U}_1 $(\%) & 6.33 & 5.11 & 5.75 & 5.92 & 5.27 \\
$\varepsilon^{N}_1 $(\%) & 0.63 & 1.22 & 0.73 & 1.08 & 1.72 \\
$\rho^{B}_1 $ & 0.08 & 0.20 & 0.08 & 1.27 & 0.29 \\
$\rho^{B}_2 $ & 0.27 & 0.56 & 0.47 & 1.03 & 0.46 \\
$\pi_2$ & 0.262 & 0.270 & 0.257 & 0.244 & 0.268 \\ \hline
\end{tabular}%
\end{table}

\newpage
\begin{figure*}[!htb]
\centering
\includegraphics[width=0.80\textwidth]{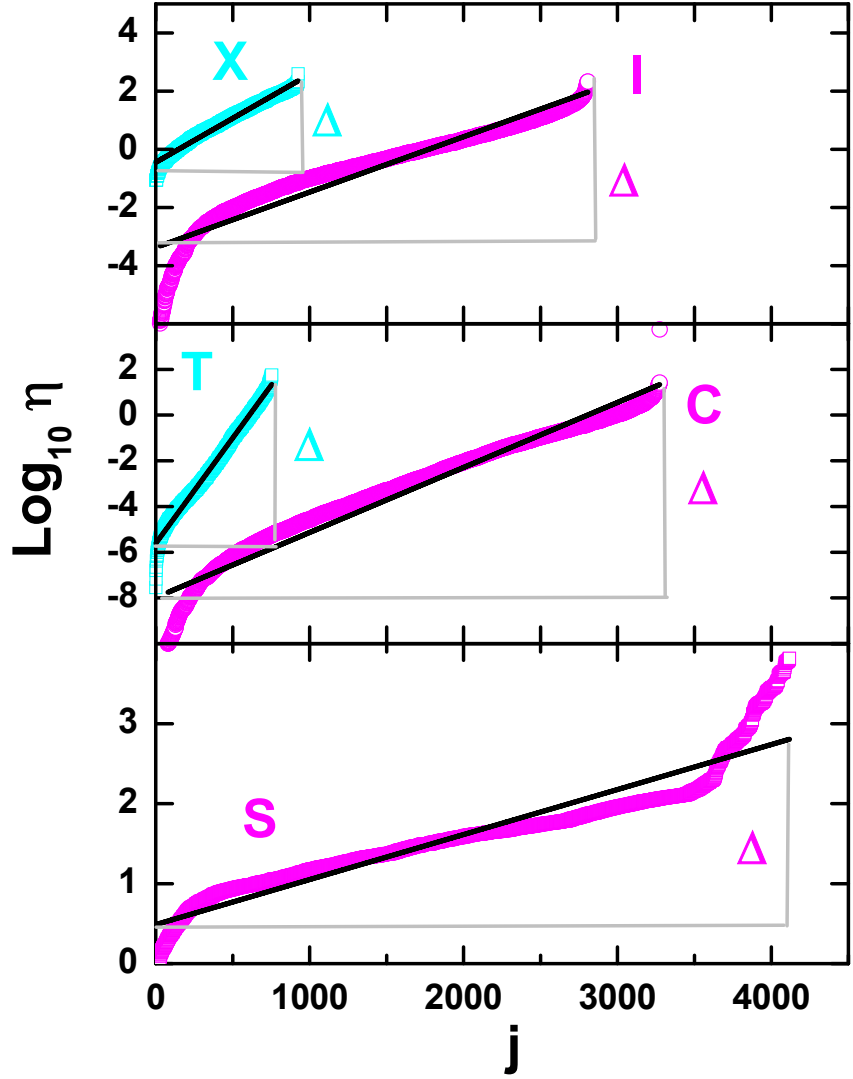}
\caption{(Color online) Symbols, logarithm of population numbers as a
function of the rank of the magnitude under study (ordered from smallest to
largest). Panels (a), (b), and (c) correspond to ionization, capture and
stopping power data sets, respectively. Solid line is the linear fitting as
given by Eq.(6) and $\Delta$ indicates the corresponding width.}
\label{Figure1}
\end{figure*}
\newpage
\begin{figure*}[!htb]
\centering
\includegraphics[width=0.80\textwidth]{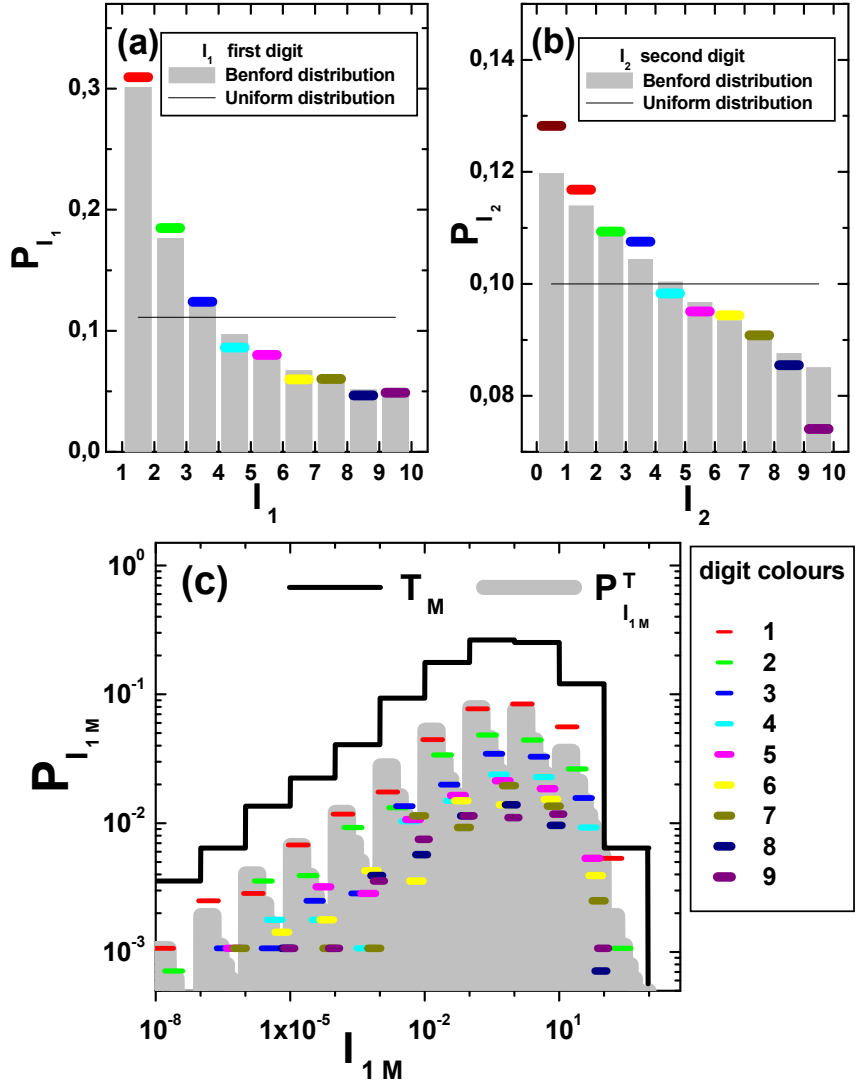}
\caption{(Color online) Color symbols are the frequencies of (a) the first
digits and (b) second digits from the data set of $nl$-ionization cross
sections. The light grey histogram represents the Bendford predictions, and
the horizontal black solid line in black denotes the uniform distribution.
(c) Color symbols are the distribution of the first digits within each
decade: $I_{1M}= I_{1}\times 10^M$. The solid line histogram represents the
top function as defined in Eq.(15). The light grey histogram represents the
Benfordization estimation given by Eq.(16).}
\label{Figure2}
\end{figure*}
\newpage
\begin{figure*}[!htb]
\centering
\includegraphics[width=0.80\textwidth]{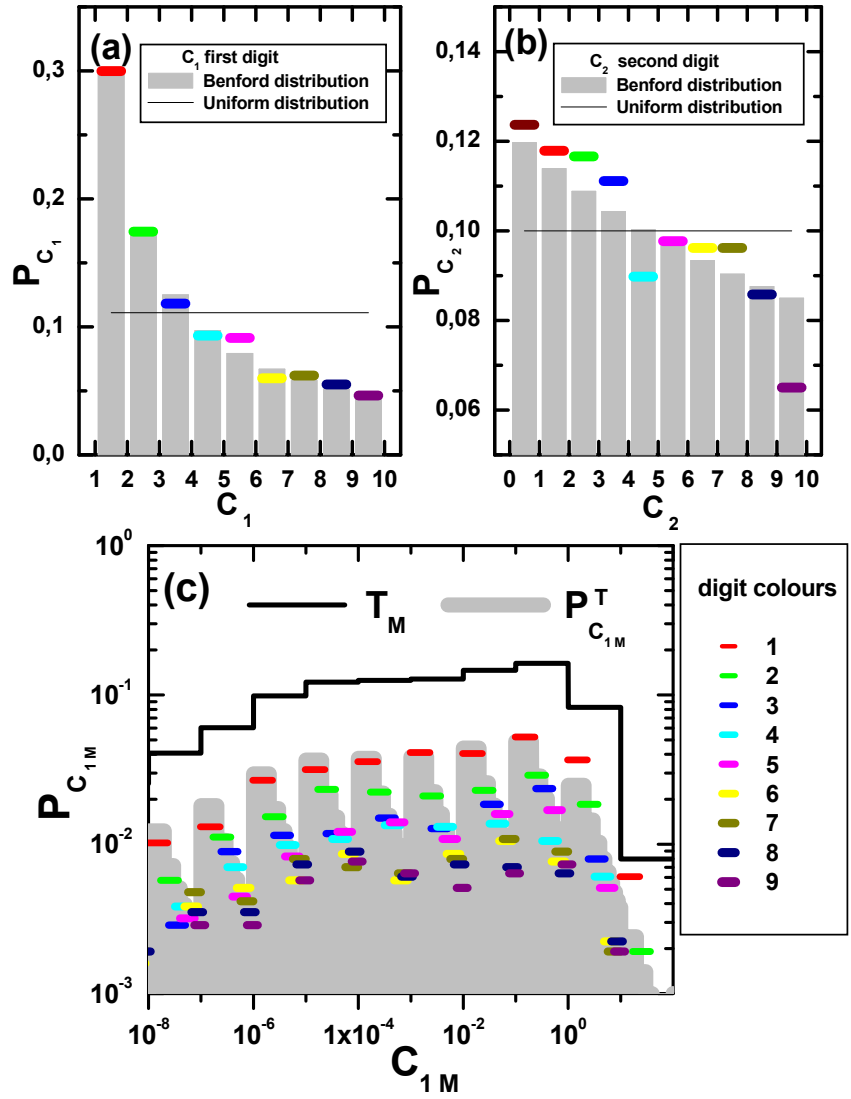}
\caption{(color on line) Similar to Fig. 2 for $nl$-capture cross sections. }
\label{Figure3}
\end{figure*}
\newpage
\begin{figure*}[!htb]
\centering
\includegraphics[width=0.80\textwidth]{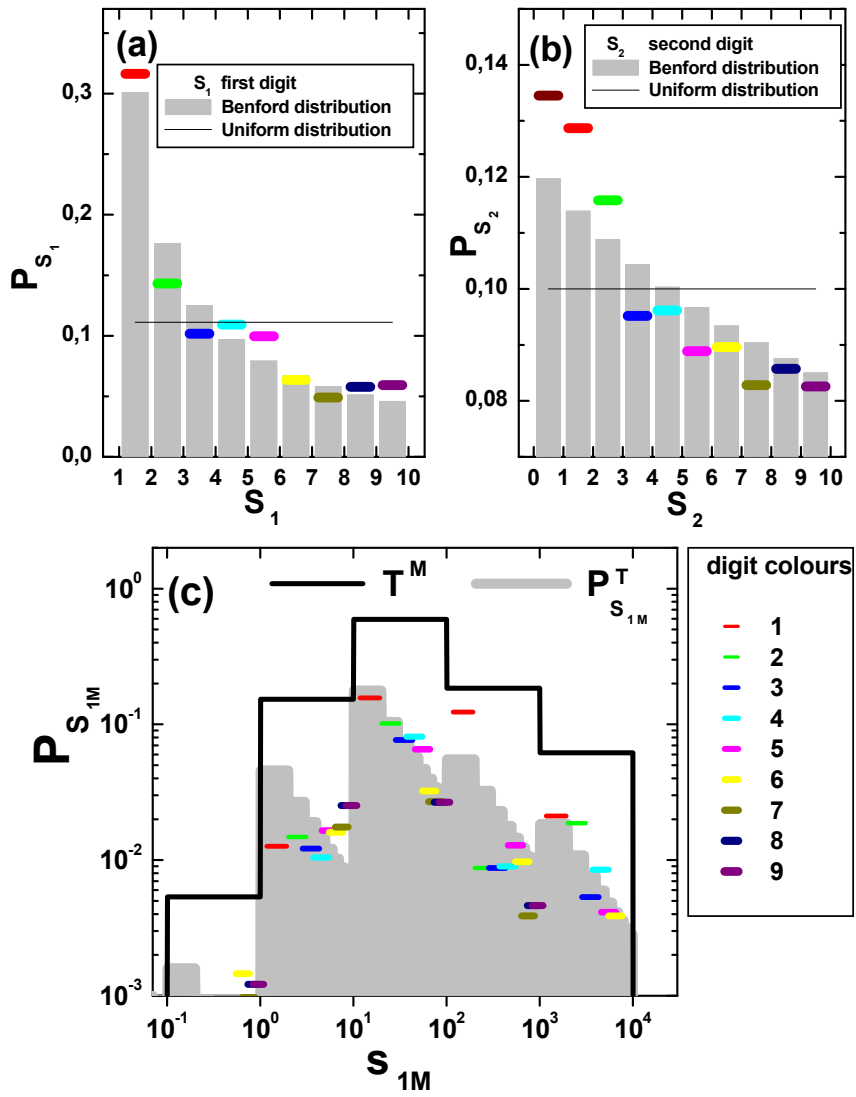}
\caption{(color on line) Similar to Fig. 2 for stopping power cross sections
from IAEA data base.}
\label{Figure4}
\end{figure*}

\end{document}